\documentclass[journal,twocolumn]{IEEEtran}
\usepackage{amsfonts}
\usepackage{times}
\usepackage{graphicx}
\usepackage{latexsym}
\usepackage{dsfont}
\usepackage{amssymb}
\usepackage{amsmath}
\usepackage{cite}
\usepackage{verbatim}
\usepackage{subfigure}

\newcommand{\figref}[1]{{Fig.}~\ref{#1}}

\def\bb0{{\mathbb{0}}}

\def\ba{{\mathbf{a}}}
\def\bb{{\mathbf{b}}}
\def\bc{{\mathbf{c}}}

\def\bff{{\mathbf{f}}}
\def\bg{{\mathbf{g}}}
\def\bh{{\mathbf{h}}}

\def\bm{{\mathbf{m}}}

\def\bp{{\mathbf{p}}}
\def\bq{{\mathbf{q}}}
\def\br{{\mathbf{r}}}

\def\bx{{\mathbf{x}}}

\def\bz{{\mathbf{z}}}
\def\b0{{\mathbf{0}}}

\def\bA{{\mathbf{A}}}
\def\bB{{\mathbf{B}}}

\def\bR{{\mathbf{R}}}

\def\bU{{\mathbf{U}}}

\def\bW{{\mathbf{W}}}

\def\bbE{{\mathbb{E}}}

\def\cA{\mathcal{A}}
\def\cB{\mathcal{B}}

\def\cF{\mathcal{F}}

\def\cN{\mathcal{N}}

\def\sf0{{\mathsf{0}}}

\usepackage{epstopdf}
\usepackage{enumerate}
\usepackage{algorithmicx}
\usepackage{algorithm}
\usepackage{amsmath}
\usepackage[noend]{algpseudocode}
\usepackage{float}
\usepackage{hyperref}
\usepackage{color}
\usepackage{makeidx}
\usepackage{bbm}
\usepackage{graphicx}

\newcommand{\sref}[1]{{Section}~\ref{#1}}

\DeclareMathOperator*{\argmax}{arg\,max}

\begin{document}
\title{Machine Learning for Reliable mmWave Systems: Blockage Prediction and Proactive Handoff}
\author{Ahmed Alkhateeb$^\dag$ and Iz Beltagy$^\ddag$\\ $^\dag$ Arizona State University, Email: alkhateeb@asu.edu\\ $^\ddag$ Allen Institute for Artificial Intelligence, Email: beltagy@allenai.org}
\maketitle

\begin{abstract}
 The sensitivity of millimeter wave (mmWave) signals to blockages is a fundamental challenge for mobile mmWave communication systems. The sudden blockage of the line-of-sight (LOS) link between the base station and the mobile user normally leads to disconnecting the communication session, which highly impacts the system reliability. Further, reconnecting the user to another LOS base station incurs high beam training overhead and critical latency problems. In this paper, we leverage machine learning tools and propose a novel solution for these reliability and latency challenges in mmWave MIMO systems. In the developed solution, the base stations learn how to predict that a certain link will experience blockage in the next few time frames using their past observations of adopted beamforming vectors. This allows the serving base station to proactively hand-over the user to another base station with a highly probable LOS link. Simulation results show that the developed deep learning based strategy successfully predicts blockage/hand-off in close to $95\%$ of the times. This reduces the probability of communication session disconnection, which ensures high reliability and low latency in mobile mmWave systems. 
\end{abstract}

\section{Introduction} \label{sec:Intro}
 
Reliability and latency are two main challenges for millimeter wave (mmWave) wireless systems \cite{MacCartney2017,Maamari2016,Alkhateeb2018}:  (i) The high sensitivity of mmWave signal propagation to blockages and the large signal-to-noise ratio gap between LOS and non-LOS links greatly affect the link reliability, and (ii) the frequent search for new base stations (BSs) after link disconnections causes critical latency overhead \cite{HeathJr2016}. This paper leverages machine learning tools to efficiently address these challenges in mobile mmWave systems.


The coordination among multiple BSs to serve the mobile user has been the main approach for enhancing the reliability of mmWave communication links \cite{MacCartney2017,Maamari2016,Alkhateeb2018}. In \cite{MacCartney2017}, extensive measurements were done for coordinated multi-point transmission at 73 GHz, and showed that simultaneously serving the user by a number of BSs noticeably improves the network coverage. This coverage performance gain was also confirmed by \cite{Maamari2016} in heterogeneous mmWave cellular networks using stochastic geometry tools. 
To overcome the large training overhead and increase the effective achievable rate in coordinated transmissions, especially for highly-mobile applications, \cite{Alkhateeb2018} proposed to use machine learning tools to predict the beamforming directions at the coordinating BSs from low-overhead features. Despite the interesting coverage gains of coordinated transmission shown in \cite{MacCartney2017,Maamari2016,Alkhateeb2018}, BSs coordination is associated with high cooperation overhead and difficult synchronization challenges.

In this paper, we develop a novel solution that enhances mmWave system reliability in high-mobile applications without requiring the high cooperation overhead of coordinated transmission. In our strategy, the serving BS uses the sequence of beams that it used to serve a mobile user over the past period of time to predict if a hand-off/blockage is going to happen in the next few moments. This allows that user and its serving BS to pro-actively hand-over the communication session to the next BS, which prevents sudden link disconnections due to blockage, improves the system reliability, and reduces the latency overhead. To do that, we develop a machine learning model based on gated recurrent neural networks that are best suited for dealing with variable-length sequences. Simulation results showed that the proposed solution predicts blockages/hand-off with almost 95$\%$ success probability and significantly improves the reliability of mmWave large antenna array systems.

\textbf{Notation}: We use the following notation: $\bA$ is a matrix, $\ba$ is a vector, $a$ is a scalar, and $\cA$ is a set. $\bA^T$, $\bA^*$ are the transpose and Hermitian (conjugate transpose) of $\bA$.  $\left[\ba\right]_n$ is the $n$th entry of $\ba$. $\bA \circ \bB$ is Hadamard product of $\bA$ and $\bB$. $\cN(\bm,\bR)$ is a complex Gaussian random vector with mean $\bm$ and covariance $\bR$.

\section{System and Channel Models} \label{sec:Model}

\begin{figure}[t]
	\centerline{
		\includegraphics[width=.9\columnwidth]{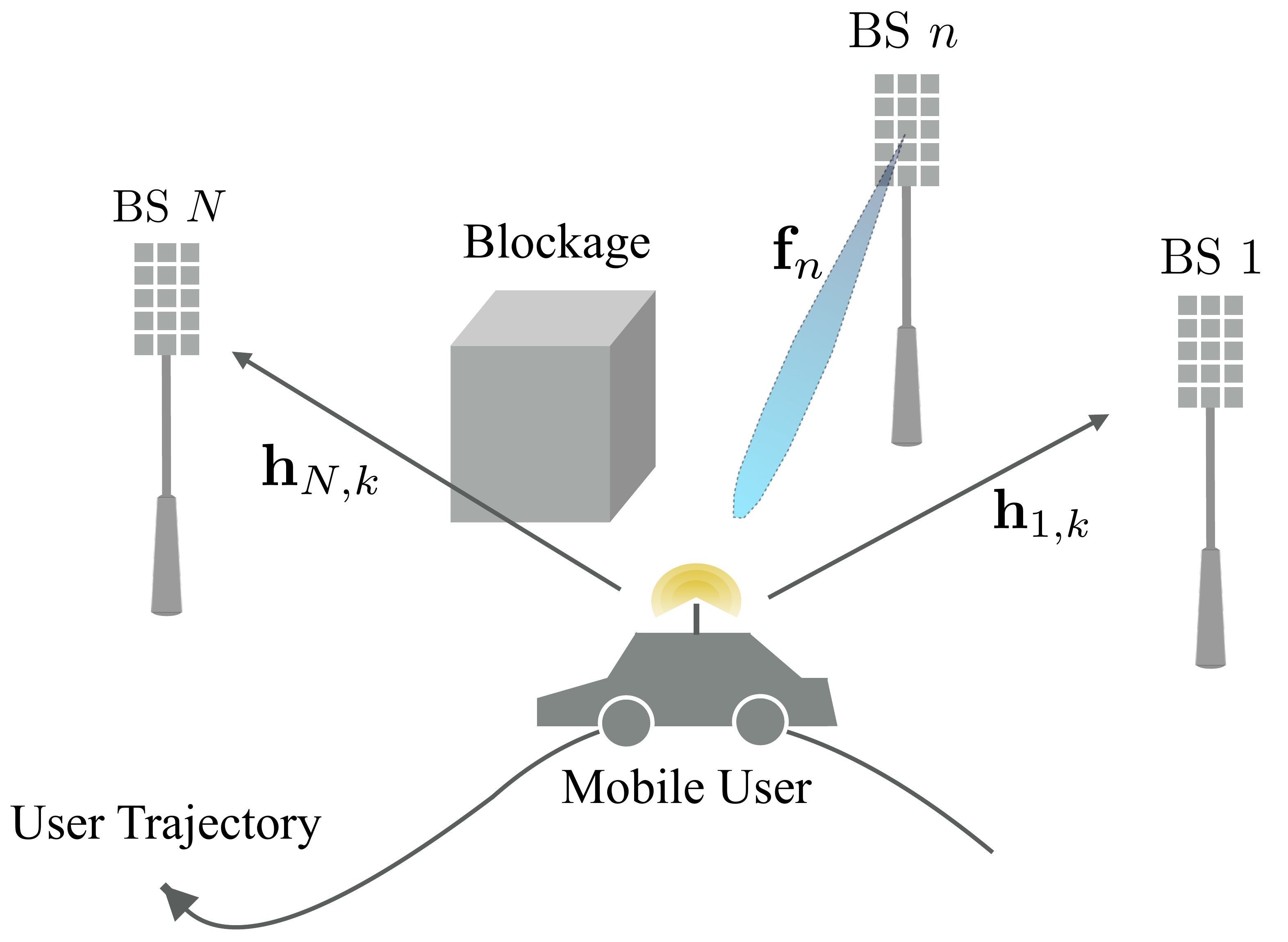}
	}
	\caption{The system model considers one user moving in a trajectory, and is served by one out of $N$ candidate BSs at every step in the trajectory.}
	\label{fig:Sys_Model}
\end{figure}

In this section, we describe the adopted mmWave system and channel models. Consider the communication setup in \figref{fig:Sys_Model}, where a mobile user is moving in a trajectory. At every step in this trajectory, the mobile user gets connected to one out of $N$ candidate base stations (BSs). For simplicity, we assume that the mobile user has a single antenna while the BS is equipped with $M$ antennas. Extending the results of this paper to the case of multi-antenna users is straight-forward. Let $\bh_{n,k}$ denote the $M \times 1$ uplink channel vector from the user to the $n$th BS at the $k$th subcarrier. If the user is connected to the $n$th BS, this BS applies a beamforming vector $\bff_n$ to serve this user. In the downlink transmission, the received signal at the mobile user on the $k$th subcarrier can then be expressed as 
\begin{equation}
y_k=\bh_{n,k}^* \bff_n s + v,
\end{equation}
where data symbol $s \in \mathbb{C}$ satisfies $\bbE\left[|s|^2\right]=P$, with $P$ the total transmit power, and $v \sim \mathcal{N}_\mathbb{C} \left(0,\sigma^2\right)$ is the receive noise at the mobile user. Due to the high cost and power consumption of mixed-signal components in mmWave large antenna array systems, beamforming processing is normally done in the analog domain using networks of phase shifters \cite{Alkhateeb2014d}. The constraints on these phase shifters limit the beamforming vectors to be selected from quantized codebooks. Therefore, we assume that the BS beamforming vector $\bff_n$ is selected from a quantized codebook $\cF$ with size/cardinality $\left|\cF\right|=M_\mathrm{CB}$. The codewords of this codebook are denoted as $\bg_m, m=1, 2, ..., M_\mathrm{CB}$. Further, we assume that the beamforming vector $\bff_n$ is selected from the codebook $\cF$ to maximize the received signal power, i.e., according to the criterion 
\begin{equation} \label{eq:BF_selection}
\bff_n = \argmax_{\bg_m \in \cF} \sum_k |\bh_{n,k}^* \bg_m|^2.
\end{equation}

We adopt a wideband geometric mmWave channel model \cite{HeathJr2016} with $L$ clusters. Each cluster $\ell, \ell=1, ..., L$ is assumed to contribute with one ray that has a time delay $\tau_\ell \in \mathbb{R}$, and azimuth/elevation angles of arrival (AoA) $\theta_\ell, \phi_\ell$. Further, let $\rho_{n}$ denote the path-loss between the user and the $n$-th BS, and $p_\mathrm{rc}(\tau)$ represents a pulse shaping function for $T_S$-spaced signaling evaluated at $\tau$ seconds. With this model, the delay-d channel  between the user and the $n$th BS follows
\begin{equation} \label{eq:d-channel}
\mathsf{\boldsymbol{h}}_{n,d} = \sqrt{\frac{M}{\rho_{n}}} \sum_{\ell=1}^L \alpha_\ell \hspace{1pt}  p(d T_\mathrm{S} - \tau_\ell) \hspace{1pt} \ba_n\left(\theta_\ell, \phi_\ell\right),
\end{equation}
where  $\ba_n\left(\theta_\ell,\phi_\ell\right)$ is the array response vector of the $n$th BS at the AoAs $\theta_\ell, \phi_\ell$.  
Given the delay-d channel in \eqref{eq:d-channel}, the frequency domain channel vector at subcarrier $k$, $\bh_{k,n}$, can be written as 
\begin{equation}
\bh_{n,k}=\sum_{d=0}^{D-1}  \mathsf{\boldsymbol{h}}_{d,n}  e^{-j \frac{2 \pi k}{K} d}.
\end{equation} 

\noindent Considering a block-fading channel model, $\left\{\bh_{k,n}\right\}_{k=1}^K$ are assumed to stay constant over the channel coherence time, denoted $T_\mathrm{C}$ \cite{Va2017a} .

\section{Problem Definition and Formulation} \label{sec:Formulation}
Maintaining good link reliability is a key challenge for mmWave communication systems, especially with mobility. This is mainly due to the high sensitivity of mmWave signals to blockages, which can frequency cause link disconnections. Further, when the link to the current BS is blocked, the mobile user incurs critical latency overhead to get connected to another BS. To overcome these challenges, can we predict that a link blockage is going to happen in the next few moments? Successful blockage prediction can be very helpful for mmWave system operation as it allows for proactive hand-off to the next BS. This proactive hand-off enhances the system reliability by ensuring session continuity and avoids the latency overhead that results from link disconnection.  In this section, we formulate the mmWave blockage prediction and proactive hand-off problem that we tackle in the next section.

\textbf{Beam sequence and hand-off status:} To formulate the problem, we first define two important quantities, namely the beam sequence and the hand-off status. Due to the user mobility, the current/serving BS needs to frequently update its beamforming vector $\bff_n \in \cF$. The frequency of updating the beams depends on a number of parameters including the user speed and the beam width. A good approximation for the period every which the BS needs to update its beam is the beam coherence time, $T_\mathrm{B}$, which is defined as \cite{Va2017a}
\begin{equation}
T_\mathrm{B}=\frac{D }{v_\mathrm{s} \sin(\alpha)} \frac{\Theta_n}{2},
\end{equation} 
where $v_\mathrm{s}$ denotes the user speed, $D$ is the distance between the user and the scatterer/reflector (or the BS in the case of LOS), $\alpha$ is the angle between the direction of travel and the direction of the main scatterer/reflector (or the BS in the case of LOS), and $\Theta_n$ defines the beam-width of the beams used by BS $n$. Now, assuming that the current/serving BS $n$ updates its beamforming vector every beam-coherence time and calling it a time step, we define $\bff_n^{(t)}$ as the beamforming vector selected by the $n$th BS to serve the mobile user in the $t$th time step, with $t=1$ representing the first time step after the handing-over to the current BS. With this, we define the beam sequence of BS $n$ until time step $t$, denoted $\cB_t$ as 
\begin{equation}
\cB_t=\left\{\bff_n^{(1)}, \bff_n^{(2)}, ..., \bff_n^{(t)} \right\}.
\end{equation}

Further, we define $s_t \in \left\{1, 2, ..., N\right\}$, as the hand-off status at the $t$th time step, with $s_t = n$ indicating the user will will stay connected to the current BS $n$ in the next time step, and $s_t \neq n$ indicating that the mobile user will hand-off to another BS in the $t+1$ time step. 
It is important to note here that predicting a hand-off in the next time step is more general that predicting a blockage, as the hand-off can happen due to a sudden blockage or a better SNR. Therefore, we will generally adopt the hand-off prediction that implicitly include link blockage prediction.

In this paper, we will focus on stationary blockages, which leaves the user mobility as the main factor of the time-varying behavior in the channel. To complete the formulation, we leverage the note that the beamforming vector that maximizes the received signal power, as defined in \eqref{eq:BF_selection}, heavily relies on the user location and the surrounding environment geometry (including blockages) \cite{Alkhateeb2018}. \textbf{Therefore, we formulate the problem as a prediction of the hand-off status at time $t+1$, i.e., $\hat{s}_t$, given the beam-sequence at time $t$, $\cB_t$}. Formally, the objective of this paper is to maximize the probability of successful blockage/hand-off prediction defined as 
\begin{equation} \label{eq:succ_prob}
\mathbb{P}\left[\hat{s}_t = s_t \left|\cB_t\right.\right].
\end{equation}

In this next section, we leverage machine learning tools to address this problem.

\section{Deep Learning Based Proactive Hand-off} 
In this section, we explain our proposed solution that uses deep learning tools, and more specifically recurrent neural networks, to efficiently predict hand-off/blockages. First, we highlight the key idea and system operation before delving into the exact machine learning modeling in \sref{subsec:ML_model}. 

\subsection{Main Idea and System Operation}
In this subsection, we briefly describe the key idea of the proposed solution as well as the learning/communications system operation. We model the problem of predicting hand-off/blockage as a sequence labeling problem. In summary, given the sequence of previous beams $\cB_t$, the serving BS predict the most likely base station that the user will connect with in the next time step. If the predicted base station is different from the current one, that indicates that a proactive hand-off needs to happen. As we mentioned in \sref{sec:Formulation}, by adopting the problem of predicting the hand-off, our system will also predict link blockages if they are going to happen in the next time step as they will require a hand-off. 

\textbf{System operation:} The proposed learning/communication system operates in two phases. In the first phase (learning), the mmWave communication system operates as normal: At every beam coherence time, the current BS will update its beamforming vector that serves the user. If the link is blocked, the user will follow the initial access process to hand-off to a new BS. During this process, the serving BS will feed the beam sequence $\cB_t$ and the hand-off status $s_t$ at every time step (beam coherence time) to its machine learning model that will use it for training. It is important to note here that these beam sequences have different lengths depending on the speed of the user, its trajectory, the time period it is spent connected to this BS, etc. As will be explained shortly, We designed our deep learning model to carefully handle this variable sequence length challenge. 

After the machine learning model is well-trained, the serving BS will leverage it to predict if the link to the user will face a blockage/hand-off in the next time-step. If a hand-off is predicted, the user and its serving BS will pro-actively initiate the hand-off process to the next BS to ensure session continuity and avoid latency problems.

\begin{figure}[t]
	\centerline{
		\includegraphics[width=.9\columnwidth]{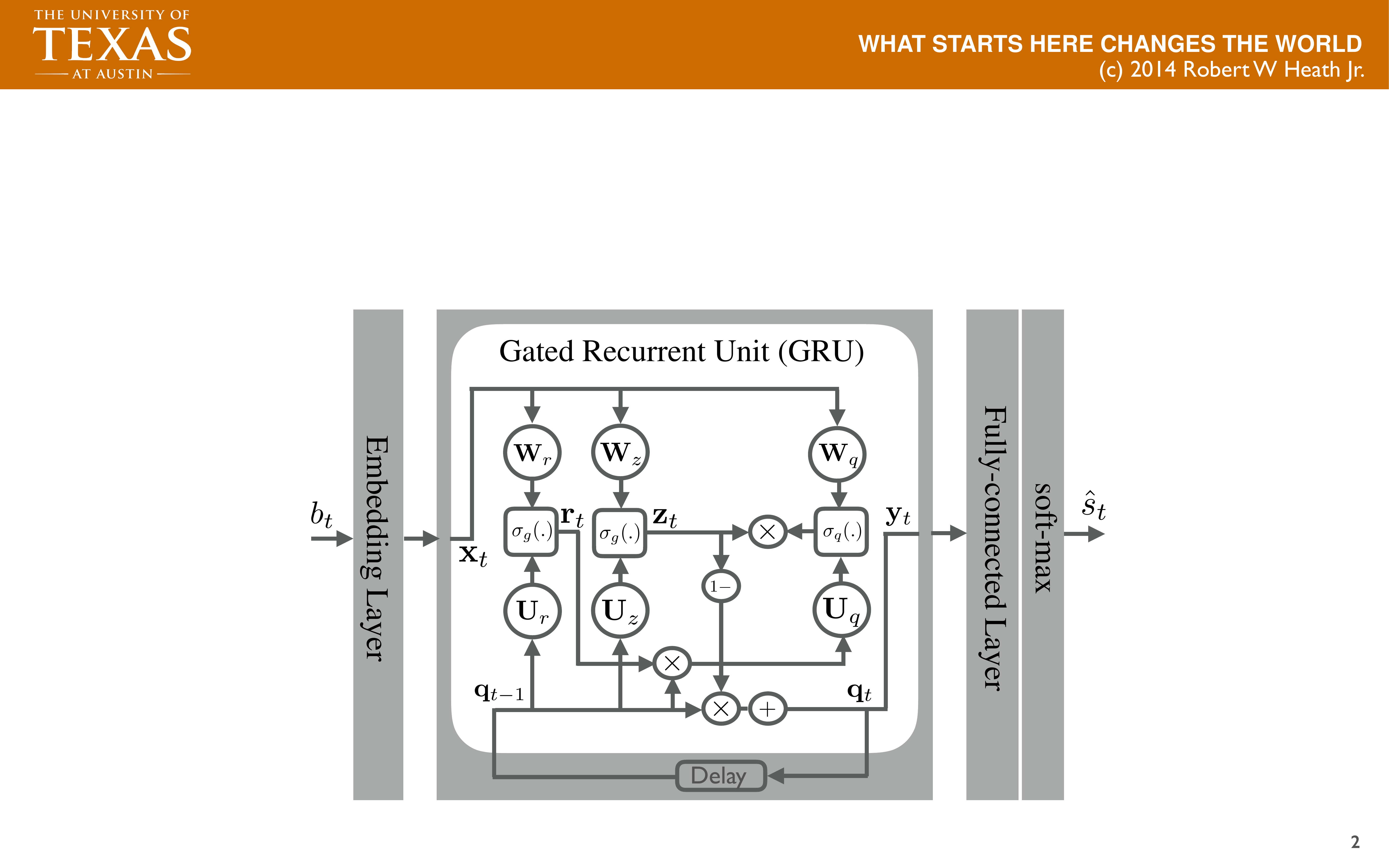}
	}
	\caption{The proposed deep learning model that leverages recurrent neural networks to predict the hand-off BS in the next time step, $\hat{s}_t$, given the past beam sequence $\cB_t$}
	\label{fig:ML_Model}
\end{figure}

\subsection{Machine Learning Model} \label{subsec:ML_model}
Next, we describe the key elements of the proposed machine learning model which is based on recurrent neural networks. 

\textbf{Input representation:} At every time step $t$, the input to our deep learning model is the $t$th beam index in the beam sequence $\cB_t$. Let $b_t \in \left\{1, 2, ..., M_\mathrm{CB}\right\}$ denote the index of the $t$th beam index, then our model starts with an embedding layer that maps every beam index $b_t$ to a vector $\bx_t$. 
\begin{align} 
\bx_t = \mathbf{embd}\left(b_t\right)
\end{align}
where $\mathbf{embd}$ is a lookup table we learn during the training. 

\textbf{Sequence processing:} The central component of our model is a Gated Recurrent Unit (GRU) \cite{Cho2014} --- a form of neural networks that is best suited for processing variable length sequnces. GRU is a recursive network that runs at every time step of the input, and it maintains a hidden state $\bq_t$ which is a function of the previous state and the current input. In addition, it has a special gating mechanism that helps with long sequences. More formally, GRU is implemented as depicted in \figref{fig:ML_Model}, and is described by the following model equations
\begin{align}
\br_t &=\sigma_g \left(\bW_r \bx_t + \bU_r \bq_{t-1} + \bc_r\right) \\
\bz_t &=\sigma_g \left(\bW_z \bx_t + \bU_z \bq_{t-1} + \bc_z\right) \\
\bq_t &= (1-\bz_t) \circ \bq_{t-1} \nonumber  \\ 
 & \hspace{10pt} + \bz_t \circ \sigma_q \left(\bW_q \bx_t + \bU_q \left(\br_t \circ \bq_{t-1} \right)+ \bc_q\right), 
\end{align}
where $\bx_t$ is the input vector, $\bq_t$ is the hidden state, $\br_t$ and $\bz_t$ are "gates" that help the model learn from long distance dependencies if needed. The weight matrices $\bW_r$, $\bW_z, \bU_r, \bU_z$ and the bias vectors $\bc_r, \bc_z, \bc_q$ are model parameters that are learned during the machine learning training phase. Finally, $\sigma_g$ and  $\sigma_q$ are nonlinear activation functions, the first is $\mathsf{sigmoid}$ and the second is $\mathsf{tanh}$. Using non linear activation functions allows the model to represent complex non-linear functions.

\textbf{Output:} At every time step $t$, the model has a hidden state $\bq_t$ which encompasses what the model learned about the sequence of beams until time step $t$, and the next step is to use $\bq_t$ to predict the most likely base station $\hat{s}_t$ for the next time step. For this, we use a fully connected layer with output size $N$ equals number of possible base stations, followed by a $\mathsf{softmax}$ activation that normalizes the  outputs into probabilities (they sum up to 1). The predicted hand-off BS (the final output of the model) can then be expressed as
\begin{align} 
\hat{s}_t = \argmax_{n \in \left\{1,2, ..., N\right\}} \mathsf{softmax}\left(\bW_\mathrm{f} \bq_t + \bc_\mathrm{f} \right)_n 
\end{align}
where $\bW_\mathrm{f}$ and $\bc_\mathrm{f}$ are the weights and biases of the fully-connected layer. The $\mathsf{softmax} (.)_n$ function takes a vector of length $N$ and outputs the probability that the $n$th BS is going to be the BS of the next time step $t+1$. The $\mathsf{softmax} (.)_n$ function is defined as follows: 
\begin{align} 
\mathsf{softmax}(\ba)_n = \frac{e^{\left[\ba\right]_n}}{\sum^{N}_{d=1}{e^{\left[\ba\right]_d}}}.
\end{align}

\begin{figure}[t]
	\centerline{
		\includegraphics[width=1\columnwidth]{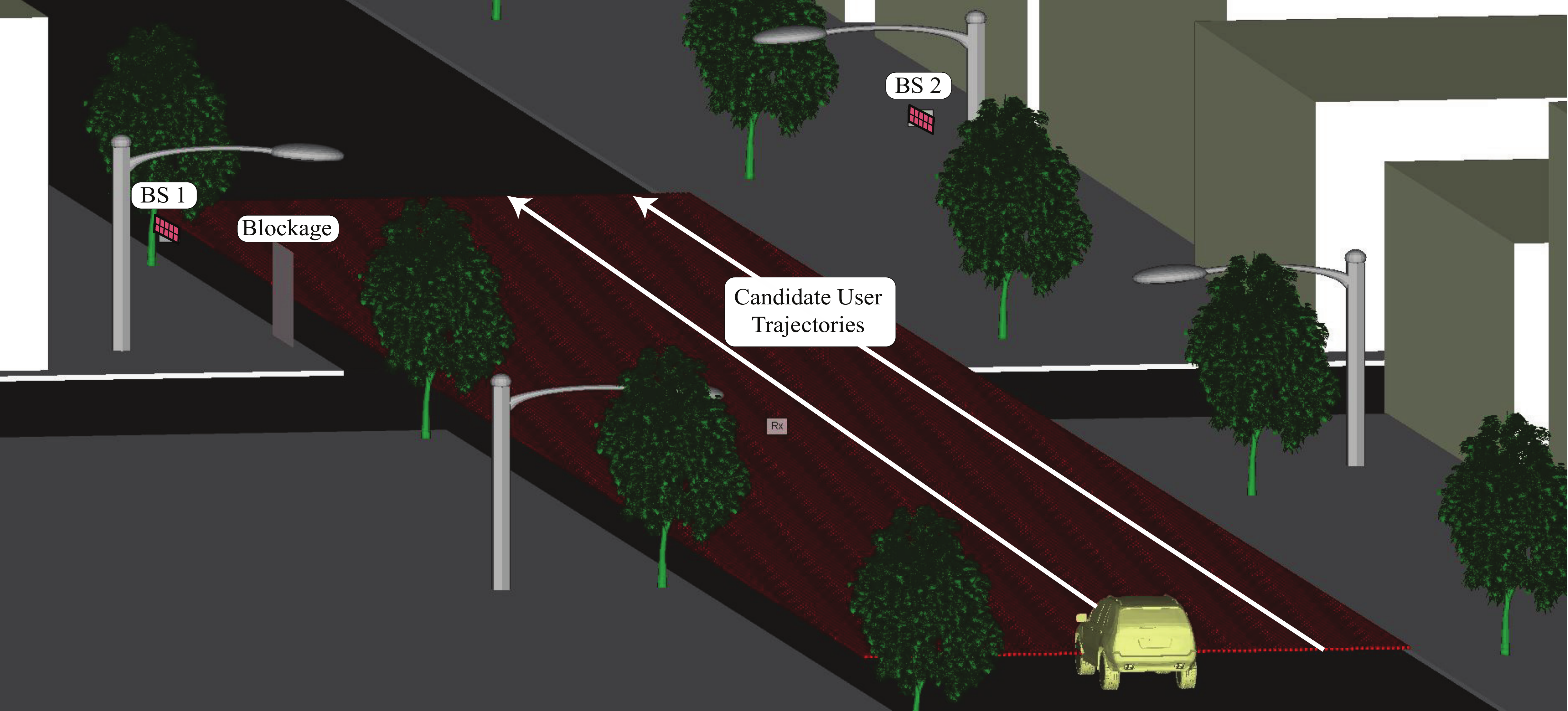}
	}
	\caption{This figure illustrates the considered simulation setup where two candidate BSs, each has ULA, serve one vehicle moving in a street.}
	\label{fig:scenario}
\end{figure}

\textbf{Training:}
The training objective in our supervised learning model is minimizing the cross entropy loss between our model predictions $\hat{s}_t$ and the actual base station of the following time step $s_t$. We compute this loss at every time step $t$ then sum over all time steps. The cross entropy loss at every time step $t$ is computed as follows: 
\begin{align} 
\mathrm{loss}(\hat{s}_t, s_t) = -\sum_i{\left[\bp\right]_{i,t} \log{\left[\hat{\bp}\right]_{i,t}}}.
\end{align}
where the reference prediction vector $\bp$ has $1$ in the entry corresponding to the index of the correct BS in $s_t$, and zero otherwise. Further, the model prediction vector $\hat{\bp}$ has the $d$th entry equals to $\mathsf{softmax}(\ba)_d, \forall d$.  

\noindent The final goal of training the model is to find the parameters $\mathbf{embd}, \bW_r, \bW_z, \bW_h, \bU_r, \bU_z, \bU_h, \bW_\mathrm{f}, \bc_r, \bc_z, \bc_h, \bc_\mathrm{f}$ that minimize this loss for all training instances.

\section{Simulation Results}

In this section, we evaluate the performance of the proposed deep-learning based proactive hand-off solution. 

\textbf{Simulation setup:} We adopt the mmWave system and channel models in \sref{sec:Model}, with two candidate BSs to serve one vehicle/mobile user moving in a street, as depicted in \figref{fig:scenario}. To generate realistic data for the channel parameters (AoAs/delay/etc.), we use the commercial ray-tracing simulator, Wireless InSite \cite{Remcom}, which is widely used in mmWave research \cite{Va2017a,Li2015a}, and is verified with channel measurements \cite{Li2015a}. Each BS is installed on one lamp post at height 4 m, and employs a 32-element uniform linear array (ULA) facing the street. The mobile user is moving in straight-line trajectories in the street, that can be any where of the street width, and with maximum trajectory length of 160 m. The trajectory starting point in randomly selected from the first 40m of the street, and the user speed is randomly selected from $\left\{8, 16, 24, 32, 40\right\}$ km/hr. The BS selects its beamforming vector from a uniformly quantized beamsteering codebook with an oversampling factor of $4$, i.e. the codebook size is $M_\mathrm{CB}=128$. At every beam coherence time, the BS beamforming vector is updated to maximize the receive SNR at the user.  During the uplink training, the MS is assumed to use $30$dBm transmit power, and the noise variance corresponds to $1$GHz system bandwidth. The system is assumed to operate at $60$GHz carrier frequency. 

\begin{figure}[t]
	\centerline{
		\includegraphics[trim={21pt 13pt 13pt 26pt},clip, width=1\columnwidth]{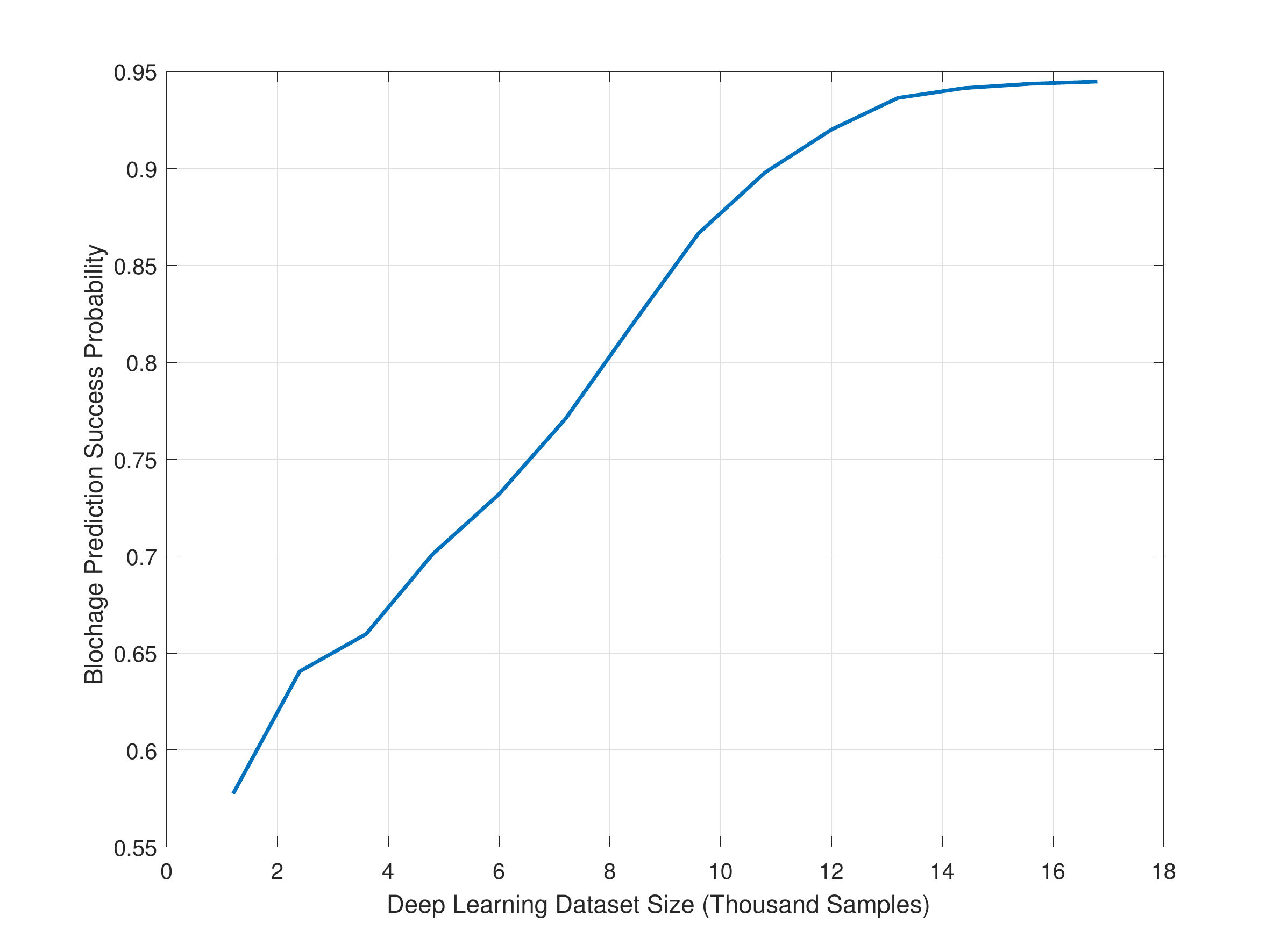}
	}
	\caption{This figure shows that the proposed deep-learning proactive hand-off solution successfully predicts blockages/hand-off with high probabilities when the machine learning model is well-trained.}
	\label{fig:results}
\end{figure}

We consider the deep learning model described in \sref{subsec:ML_model}. The neural network model has an embedding that outputs vectors of length $20$ to the GRU unit, with maximum sequence length of $454$. Since we have only 2 BSs in out experiment, the fully-connected layer has only two outputs that go to the softmax function. We use the Adam optimizer \cite{Kingma2014}. In the deep learning experimental work, we used the \emph{Keras} libraries \cite{Branchaud-Charron} with a \emph{TensorFlow} backend.

\textbf{Hand-off/Blockage prediction:} To evaluate the performance of the proposed deep-learning based proactive hand-off solution, \figref{fig:results} plots the blockage/hand-off successful prediction probability, defined in \eqref{eq:succ_prob}, versus the training size. \figref{fig:results} shows that with sufficient dataset size (larger than 12 thousand samples), the machine learning model successfully predicts the hand-off with more than $90 \%$ probability, given only the sequence of past beams $\cB_t$. This illustrates the potential of the proposed solution in enhancing the reliability of next-generation mmWave systems.

\section{Conclusion}
In this paper, we proposed a novel solution for the reliability and latency problem in mobile mmWave systems. Our solution leveraged deep learning tools to efficiently predict blockage and the need for hand-off. This allows the mobile user to proactively hand-off to the next BS without disconnecting the session or suffering from high latency overhead due to sudden link disconnections. The simulation results showed that the developed proactive hand-off strategy can successfully predicts blockage/hand-off with high probability when the machine learning model is trained with reasonable dataset sizes. In the future, it is interesting to extend the developed solution to multi-user systems and account for mobile blockages.

\bibliographystyle{IEEEtran}

\end{document}